\documentclass[10pt,english,aps,showkeys,showpacs,nofootinbib]{revtex4-1}
\usepackage[T1]{fontenc}
\usepackage{textcomp}
\usepackage[utf8]{inputenc}
\usepackage{geometry}
\usepackage{color}
\usepackage{babel}
\usepackage{amsmath}
\usepackage{amssymb}
\usepackage{stackrel}
\usepackage[bookmarks=false,
 breaklinks=false,pdfborder={0 0 1},backref=section,colorlinks=false]
 {hyperref}
\begin{document}
\selectlanguage{english}
\title{Behavior of the Feshbach-Villars oscillator (FVO) in Gürses space-time
under Coulomb-type potential.}
\author{Abdelmalek Bouzenada }
\email{ abdelmalekbouzenada@gmail.com; abdelmalek.bouzenada@univ-tebessa.dz}

\affiliation{Laboratory of theoretical and applied Physics,~~\\
 Echahid Cheikh Larbi Tebessi University, Algeria}
\author{Abdelmalek Boumali}
\email{boumali.abdelmalek@gmail.com}

\affiliation{Laboratory of theoretical and applied Physics,~~\\
 Echahid Cheikh Larbi Tebessi University, Algeria}
\author{Omar Mustafa }
\email{omar.mustafa@emu.edu.tr }

\affiliation{Department of Physics, Eastern Mediterranean University, ~~\\
 G. Magusa, north Cyprus, Mersin 10 - Turkiye}
\author{Hassan Hassanabadi }
\email{h.hasanabadi@shahroodut.ac.ir }

\affiliation{Department of Physics, Shahrood University, ~~\\
 Shahrood, P.O. Box 3619995161-316, Iran.}
\date{\today}
\begin{abstract}
\textcolor{black}{Our research aims to investigate how the gravitational
field influences the spectroscopic structure of the Feshbach-Villars
oscillator in Gürses space-time. To achieve this, we utilize the first-order
Feshbach-Villars version of the Klein-Gordon equation, which is a
relativistic wave equation for spinless particles. We examine the
oscillator's quantum mechanical behavior in the presence of a Coulomb-type
potential, and calculate the resulting wave functions and energy levels
for both free and interacting scenarios. In addition, we study the
interaction between the Coulomb-type potential and Gürses space-time
affects the Feshbach-Villars oscillator's behavior, specifically with
regard to its spectroscopic structure. This study has important implications
for our understanding of the interplay between quantum mechanics,
relativity, and gravitational fields at the microscopic level. } 
\end{abstract}
\keywords{Klein-Gordon equation, Feshbach--Villars Oscillator, topological
defects, Gürses space-time, coulomb-type potential, Biconfluent Heun
function .}
\pacs{04.62.+v; 04.40.\textminus b; 04.20.Gz; 04.20.Jb; 04.20.\textminus q;
03.65.Pm; 03.50.\textminus z; 03.65.Ge; 03.65.\textminus w; 05.70.Ce}
\maketitle

\section{Introduction }

The study of how the gravitational field affects the dynamics of quantum
mechanical systems is of great interest. General relativity, which
explains gravity (GR)\citep{key-1} as a geometric property of space-time,
provides a convincing explanation for the classical gravitational
field as a manifestation of space-time curvature and predicts phenomena
like gravitational waves \citep{key-2} and black holes\citep{key-3}.
On the other hand, quantum mechanics (QM) is a framework for understanding
the behavior of particles on a small scale and describing interactions
among fundamental forces\citep{key-4}. However, developing a unified
theory of quantum gravity\citep{key-5} that reconciles general relativity
and quantum mechanics has proven difficult due to various unresolved
technical challenges and roadblocks \Citep{key-6,key-7} .

A method to gain a general understanding of how the gravitational
field affects relativistic particles at the quantum level is to apply
principles of relativistic dynamics of particles in flat Minkowski
space to a curved background geometry\citep{key-8,key-9}. This approach
can be applied to various models involving curvature, which can provide
predictions for macroscopic observables that can be experimentally
verified in astrophysics and cosmology. Additionally, it is important
to understand the thermodynamic behavior of relativistic particles
when considering gravitational effects \citep{key-10,key-11,key-12}
and to analyze the statistical quantities that underlie this behavior.
This can lead to important insights into the quantum behavior of gravity.
By comprehending these features, scientists can obtain essential results
that will contribute to our understanding of the complex phenomena
in the universe.

The study of topological defects, including domain walls, cosmic strings,
monopoles, and textures, has been a highly active field of research
in condensed matter physics, cosmology, astrophysics, and elementary
particle models for several decades. These structures are believed
to have emerged due to the Kibble mechanism \citep{key-13} during
the cooling of the early universe in symmetry-breaking phase transitions\Citep{key-14,key-15}.
Among these structures \citep{key-16,key-17}, cosmic strings have
received particular attention (see \Citep*{key-18} for more information),
and their static or rotating presence can have noticeable effects,
such as seeding galaxy formation and causing gravitational lensing.
By examining cosmic strings and their properties, we can gain insight
into particle physics at extremely high energies in various settings.
Moreover, the idea that cosmic strings could act as superconducting
wires has been proposed in current physics, which would have significant
implications.

The quantum harmonic oscillator (HO) is a highly significant tool
in various areas of theoretical physics \Citep{key-19}. It is a model
that has been thoroughly investigated and can be employed to explore
various challenging issues in quantum mechanics \Citep{key-20}. Additionally,
the relativistic version of the HO serves as a valuable model for
understanding molecular, atomic, and nuclear interactions. Its ability
to provide a complete set of exact analytical solutions can result
in fundamentally different interpretations of mathematical and physical
phenomena, which, in turn, can lead to a wide range of applications
in related fields.

The Dirac oscillator (DO) plays a crucial role in the behavior of
various relativistic quantum systems. It has been shown by Itô et
al \citep{key-21}. that the dynamics of spin-1/2 particles with a
linear trajectory are heavily influenced by this system. The non-relativistic
limit of the DO is an ordinary harmonic oscillator with a considerable
spin-orbit coupling term. Furthermore, Moshinsky and Szczepaniak \citep{key-22}
demonstrated that the DO can be derived from the free Dirac equation
by introducing an external linear potential through \textcolor{black}{the
replacement of the regular radial momentum operator $\hat{p}_{r}$
with its non-minimal coupling form $\hat{p}_{r}\rightarrow\hat{p}_{r}-i\eta r$
, where $\eta=m\omega$, recovers Moshinsky and Szczepaniak Dirac
oscillator \citep{key-22}, Such a non-minimal coupling from can,
in fact, be generalized to its position-dependent mass (PDM) counter
part $\hat{p}_{r}\rightarrow\hat{p}_{r}-i\mathcal{M}_{r}$, where
$\mathcal{M}_{r}=\frac{f(r)'}{4f(r)}$, which is inherited form PDM-Schrodinger
particles\citep{key-23,key-24,key-25} and PDM Klein-Gordon (KG) particles
\citep{key-26,key-27,key-28,key-29,key-30}.}

\textcolor{black}{Inspired by Moshinsky and Szczepaniak's Dirac oscillators
\citep{key-22}}, the development of the Klein-Gordon oscillator (KGO)
was influenced by the Dirac oscillator (DO) and is a similar formalism
for bosonic particles \citep{key-31,key-32}. Recently, researchers
have been exploring the covariant version of this model in various
configurations, including curved space-times. The subject of relativistic
quantum motion of scalar and vector particles under gravitational
effects caused by different curved space-time geometries has received
significant attention. For example, Ref.\Citep{key-33} examines the
interaction between KGO and topological defects in Kaluza-Klein theory.
Ref.\Citep{key-34} investigates the relativistic quantum dynamics
of spin-0 particles in a spinning cosmic string space-time with Coulomb-type
scalar and vector potentials. Ref. \Citep{key-35} explores the rotating
effects on the scalar field in cosmic string space-time, space-time
with space-like dislocation, and space-time with spiral dislocation.
Recently, the authors of Ref.\Citep{key-36} studied the KGO in a
cosmic string space-time and examined the effects of the rotating
frame and non-commutativity in momentum space. Additionally, Ref.\Citep{key-37}
exposes the KGO to a magnetic quantum flux in the presence of a Cornell-type
scalar and Coulomb-type vector potentials in a spinning cosmic string
space-time.

Various researchers have attempted to study the time evolution and
wave functions of spin-0 and spin-1 bosons, as well as spin-1/2 fermions,
in a relativistic context using Hamiltonian equations, such as the
Schrodinger equation\Citep{key-37,key-38,key-39}. One of the methods
used in this regard is the Feshbach-Villars (FV) equations\Citep{key-40},
which were developed to allow for a relativistic interpretation of
the second-order Klein-Gordon equation for single particles. The FV
equations involve splitting the wave function into two components
to obtain an equation with a first-order time derivative. In recent
years, a significant number of studies have been devoted to exploring
the dynamical properties of single particles in a relativistic context
and solving their wave equations using the FV approach. For example
(see \Citep{key-41,key-42,key-43,key-44,key-45,key-46,key-47}), Bouzenada
et al. \citep{key-48} investigated the Feshbach-Villars oscillator
(FVO) case in a spinning cosmic string space-time and reported their
findings on the thermal properties and density of the system.

The present study examines the influence of a Coulomb-like potential
on the Klein-Gordon oscillator. In recent publications \citep{key-49,key-50,key-51,key-52,key-53}
, confinement of a relativistic scalar particle to a Coulomb potential
has been investigated \citep{key-52}. One method of incorporating
a scalar potential into the Klein-Gordon equation is by introducing
the electromagnetic 4-vector potential in the same way as the momentum
operator, i.e., $p_{\alpha}=i\partial_{\alpha}$ is modified as $p_{\alpha}\rightarrow p_{\alpha}-qA_{\alpha}\left(x\right)$.
Another approach, as proposed in Ref.\citep{key-53}, is to modify
the mass term as $m\rightarrow m+S(\overrightarrow{r},t)$, where
$S(\overrightarrow{r},t)$ represents the Lorentz scalar potential.
This modification in the mass term has been studied in the context
of a Dirac particle in the presence of a static scalar and Coulomb
potential \citep{key-53}, and a relativistic scalar particle in cosmic
string spacetime \citep{key-54}. In this work, we investigate the
impact of a Coulomb-like potential on the Klein-Gordon oscillator
by introducing the scalar potential as a modification to the mass
component of the Klein-Gordon equation. We obtain solutions to the
bound state Klein-Gordon equation for both attractive and repulsive
Coulomb-like potentials and reveal a quantum effect where the angular
frequency of the Klein-Gordon oscillator depends on the quantum numbers
of the system, indicating that not all angular frequency values are
possible.

This article is structured in the following manner. In the upcoming
section, we will examine the FV representation of Feshbach-Villars
(Spin-0) in Minkowski Space-time. Following that, we will introduce
the FV representation of Spin-0 Particle in Gürses Space-time and
PDM of FV oscillator in Gürses space-time in the subsequent section.
The next section will focus on investigating the solution of Free
equation and oscillator in Gürses space-time. Subsequently, we will
explore the interaction with Coulomb-type potential for the free and
oscillator in this background in the following section. Finally, we
will provide concluding remarks. It should be noted that natural units
$\hbar=c=1$ will be consistently employed throughout this article,
and our metric convention is $(+,-,-,-)$.

\section{The FV Representation of Feshbach-Villars (Spin-0) in Minkowski Space-time}

\subsection{An Overview of the Feshbach-Villars Approximation}

This section discusses the relativistic quantum description of a spin-0
particle propagating in Minkowski space-time using the metric tensor
${\color{black}\eta_{\mu\nu}=\text{diag}}\mathinner{\color{black}\left(1,-1,-1,-1\right)}$.The
usual covariant KG equation for a scalar massive particle $\Phi$
with mass $m>0$ is \Citep{key-55,key-56} 
\begin{equation}
\left(\eta^{\mu\nu}D_{\mu}D_{\nu}+m^{2}\right)\Phi(x,t)=0,\label{eq:01}
\end{equation}
The minimally-coupled covariant derivative is denoted by ${\color{black}D_{\mu}=}\mathinner{\color{black}\left(\partial_{\mu}-ieA_{\mu}\right)}$.
The classical four momentum is $p_{\mu}=\left(E,-p_{i}\right)$, while
the electromagnetic four potential is $A_{\mu}=\left(A_{0},-A_{i}\right)$.
The magnitude of the particle charge is given by e.

It is worth noting at this point that \eqref{eq:01} may be expressed
in Hamiltonian form with the time first derivative, i.e. as a Schrödinger-type
equation. 
\begin{equation}
\mathcal{H}\Phi\left(x,t\right)=i\,\frac{\partial}{\partial t}\Phi(x,t),\label{eq:02}
\end{equation}
The Hamiltonian $\mathcal{H}$ may be defined using the FV linearization
process, which involves converting \ref{eq:01} to a first order in
time differential equation. The two component wave function \Citep{key-53}
is introduced. \Citep{key-57},
\begin{equation}
\Phi(x,t)=\left(\begin{array}{c}
\phi_{1}(x,t)\\
\phi_{2}(x,t)
\end{array}\right)=\frac{1}{\sqrt{2}}\left(\begin{array}{c}
1+\frac{i}{m}\mathcal{D}\\
1-\frac{i}{m}\mathcal{D}
\end{array}\right)\psi(x,t),\label{eq:03}
\end{equation}
Here, $\psi(x,t)$ obeys the KG wave equation, and $\mathcal{D}$
is defined in such a way that 
\begin{equation}
\mathcal{D}=\frac{\partial}{\partial t}+ieA_{0}(x),\label{eq:04}
\end{equation}
The aforementioned transformation \eqref{eq:03} involves inserting
wave functions that meet the requirements. 
\begin{equation}
\psi=\phi_{1}+\phi_{2},\qquad i\mathcal{D}\psi=m\left(\phi_{1}-\phi_{2}\right).\label{eq:05}
\end{equation}
It is more convenient to write for our subsequent review, 
\begin{equation}
\begin{aligned}\phi_{1} & =\frac{1}{2m}\left[m+i\frac{\partial}{\partial t}-eA_{0}\right]\psi\\
\phi_{2} & =\frac{1}{2m}\left[m-i\frac{\partial}{\partial t}+eA_{0}\right]\psi,
\end{aligned}
\label{eq:06}
\end{equation}
Eq. \eqref{eq:01} becomes equivalent 
\begin{equation}
\begin{aligned}\left[i\frac{\partial}{\partial t}-eA_{0}\right]\left(\phi_{1}+\phi_{2}\right) & =m\left(\phi_{1}-\phi_{2}\right)\\
\left[i\frac{\partial}{\partial t}-eA_{0}\right]\left(\phi_{1}-\phi_{2}\right) & =\left[\frac{\left(p_{i}-eA_{i}\right)^{2}}{m}+m\right]\left(\phi_{1}+\phi_{2}\right),
\end{aligned}
\label{eq:07}
\end{equation}
The addition and subtraction of these two equations yields a system
of first order coupled differential equations 
\begin{equation}
\begin{aligned}\frac{\left(p_{i}-eA_{i}\right)^{2}}{2m}\left(\phi_{1}+\phi_{2}\right)+\left(m+eA_{0}\right)\phi_{1} & =i\frac{\partial\phi_{1}}{\partial t}\\
\frac{-\left(p_{i}-eA_{i}\right)^{2}}{2m}\left(\phi_{1}+\phi_{2}\right)-\left(m-eA_{0}\right)\phi_{2} & =i\frac{\partial\phi_{2}}{\partial t},
\end{aligned}
\label{eq:08}
\end{equation}
The FV Hamiltonian of a scalar particle in the presence of electromagnetic
interaction may be expressed using Eqs. \eqref{eq:08} as
\begin{equation}
\mathcal{H}_{K-G}=\left(\tau_{3}+i\tau_{2}\right)\frac{\left(p_{i}-eA_{i}\right)^{2}}{2m}+m\tau_{3}+eA_{0}(x),\label{eq:09}
\end{equation}
where $\tau_{i}\,\left(i=1,2,3\right)$ are the conventional $2\times2$
Pauli matrices given by 
\begin{equation}
\tau_{1}=\left(\begin{array}{cc}
0 & 1\\
1 & 0
\end{array}\right),\quad\tau_{2}=\left(\begin{array}{cc}
0 & -i\\
i & 0
\end{array}\right),\quad\tau_{3}=\left(\begin{array}{cc}
1 & 0\\
0 & -1
\end{array}\right).\label{eq:10}
\end{equation}
It's worth noting that the Hamiltonian \eqref{eq:09} meets the generalized
hermiticity requirement (If there is an inevitable, Hermitian, linear
operator $\beta$ such that $\mathcal{H}^{\dagger}=\beta\mathcal{H}\beta^{-1}$,
the Hamiltonian $\mathcal{H}$ is said to be pseudo-Hermitian. \Citep{key-58}).
\begin{equation}
\mathcal{H}_{K-G}=\tau_{3}\left(\mathcal{H}_{K-G}^{\dagger}\right)\tau_{3},\qquad\mathcal{H}_{K-G}^{\dagger}=\tau_{3}\left(\mathcal{H}_{K-G}\right)\tau_{3}.\label{eq:11}
\end{equation}
The one dimensional FV Hamiltonian reduces to for free particle propagation,
i.e., no interaction is assumed left $\left(A_{\mu}=0\right)$. 
\begin{equation}
\mathcal{H}_{0}=\left(\tau_{3}+i\tau_{2}\right)\frac{p_{x}^{2}}{2m}+m\tau_{3},\label{eq:12}
\end{equation}
The solutions to the time-independent free Hamiltonian are simply
stationary states. Assuming the solution \Citep{key-36}, 
\begin{equation}
\Phi\left(x,t\right)=\Phi\left(x\right)e^{-iEt}=\left(\begin{array}{c}
\phi_{1}\left(x\right)\\
\phi_{2}\left(x\right)
\end{array}\right)e^{-iEt},\label{eq:13}
\end{equation}
with E denoting the system's energy. As a result, Eq. \eqref{eq:02}
may be represented as 
\begin{equation}
\mathcal{H}_{0}\Phi\left(x\right)=E\Phi\left(x\right),\label{eq:14}
\end{equation}
This is the one-dimensional FV equation of the free relativistic spin-0
particle, and it is performed in order to have an alternate Schrödinger-type
to KG equation. In what follows, the aforesaid approach will be utilized
to determine the Gürses solutions to wave equations in curved space-time.

\section{The FV Representation of Spin-0 Particle in Gürses Space-time}

The goal of this part is to investigate the KGO in the background
geometry of a Gürses space-time using the FV technique. It is widely
known that the generally covariant relativistic wave equations of
a scalar particle in a Riemannian space-time characterized by the
metric tensor $g_{\mu\nu}$ may be found by reformulating the KG equation
so that( see, e.g, the textbooks\citep{key-8,key-9} ) 
\begin{equation}
\left(\square+m^{2}-\mathcal{\xi}R\right)\Phi(x,t)=0,\label{eq:15}
\end{equation}
where $\square$ is the Laplace-Beltrami operator denoted by 
\begin{equation}
\square=g^{\mu\nu}D_{\mu}D_{\nu}=\frac{1}{\sqrt{-g}}\partial_{\mu}\left(\sqrt{-g}g^{\mu\nu}\partial_{\nu}\right),\label{eq:16}
\end{equation}
$\xi$ denotes a real dimensionless coupling constant, and R is the
Ricci scalar curvature given by $R=g^{\mu\nu}R_{\mu\nu}$, where $R_{\mu\nu}$
is the Ricci curvature tensor. The inverse metric tensor is $g^{\mu\nu}$,
and $g=\det\left(g_{\mu\nu}\right)$.

We would now want to investigate the quantum dynamics of spin-0 particles
in the space-time caused by a (1+2)-dimensional , as well as develop
the relevant FV formulation.

\subsection{Feshbach-Villars oscillators in Gürses space-time}

Note that this section closely follows the works of Silenko \citet{key-67},
where the metric signature used is ${\color{black}\eta_{\mu\nu}=\text{diag}(1,-1,-1}\mathclose{\color{black})}$
and consequently $\sqrt{-g}$ below follows this signature with
\begin{equation}
\mathcal{Y}=\frac{1}{2}\left\{ \partial_{i},\sqrt{-g}\frac{g^{0i}}{g^{00}}\right\} .\label{eq:20}
\end{equation}
The anti-commutator is denoted by the curly bracket in Eq. \eqref{eq:20}.
The Hamiltonian for the aforementioned transformation is 
\begin{equation}
\mathcal{H}_{GFVT}=\tau_{z}\left(\frac{\mathcal{N}^{2}+\mathcal{T}}{2\mathcal{N}}\right)+i\tau_{y}\left(\frac{-\mathcal{N}^{2}+\mathcal{T}}{2\mathcal{N}}\right)-i\mathcal{Y},\label{eq:21}
\end{equation}
with 
\begin{align}
\mathcal{T}' & =\partial_{i}\frac{G^{ij}}{g^{00}}\partial_{j}+\frac{m^{2}-\xi R}{g^{00}}+\frac{1}{\mathcal{F}}\nabla_{i}\left(\sqrt{-g}G^{ij}\right)\nabla_{j}\left(\frac{1}{\mathcal{F}}\right)+\sqrt{\frac{\sqrt{-g}}{g^{00}}}G^{ij}\nabla_{i}\nabla_{j}\left(\frac{1}{\mathcal{F}}\right)+\frac{1}{4\mathcal{F}^{4}}\left[\nabla_{i}\left(\mathcal{U}^{i}\right)\right]^{2}\nonumber \\
 & \qquad-\frac{1}{2\mathcal{F}^{2}}\nabla_{i}\left(\frac{g^{0i}}{g^{00}}\right)\nabla_{j}\left(\mathcal{U}^{i}\right)-\frac{g^{0i}}{2g^{00}\mathcal{F}^{2}}\nabla_{i}\nabla_{j}\left(\mathcal{U}^{i}\right),\label{eq:22}
\end{align}
where 
\begin{equation}
G^{ij}=g^{ij}-\frac{g^{0i}g^{0j}}{g^{00}},\qquad\mathcal{F}=\sqrt{g^{00}\sqrt{-g}},\qquad\mathcal{U}^{i}=\sqrt{-g}g^{0i}.\label{eq:23}
\end{equation}
Here and everywhere $(i,j=1,2)$. We see that the initial FV transformations
are fulfilled for $\mathcal{N}=m$.

The Gürses metric in (1+2)-dimensions exhibits both circular stationary
and rotational symmetry. 
\begin{align}
ds^{2}=g_{\mu\nu}dx^{\mu}dx^{\nu}=dt^{2}-dr^{2}+2\Omega r^{2}dtd\varphi-r^{2}\left(1-\Omega^{2}r^{2}\right)d\varphi^{2} & .\label{eq:17}
\end{align}
where, the metric signature is ${\color{black}\eta_{\mu\nu}=\text{diag}(1,-1,-1}\mathclose{\color{black})}$
and $\Omega=\pm|\Omega|$ represents positive/negative vorticities
of the rotating spacetime. Here, $r\ge0$, $0\le\varphi\le2\pi$,
The metric tensor and its inverse are, respectively, given by 
\begin{equation}
g_{\mu\nu}=\left(\begin{array}{ccc}
1 & 0 & \Omega r^{2}\\
0 & -1 & 0\\
\Omega r^{2} & 0 & -r^{2}\left(1-\left(\Omega r\right)^{2}\right)
\end{array}\right),\,g^{\mu\nu}=\left(\begin{array}{ccc}
(1-\Omega^{2}r^{2}) & 0 & \Omega\\
0 & -1 & 0\\
\Omega & 0 & -1/r^{2}
\end{array}\right),\label{eq:18}
\end{equation}
with $\text{det}(g_{\mu\nu})=-r^{2}$. Now, following the same procedure
used in Ref. \citet{key-49}, we obtain 
\begin{equation}
T=\frac{1}{g^{00}\sqrt{-g}}\partial_{i}(\sqrt{-g}\,g^{ij}\,\partial_{j})+\frac{m^{2}-\zeta R}{g^{00}}-\mathcal{Y}^{2},\label{e1}
\end{equation}
with $i=1,2$ 
\begin{equation}
\mathcal{Y}=\frac{1}{g^{00}\sqrt{-g}}\,\{\partial_{i},\sqrt{-g}g^{0i}\}=\frac{g^{02}}{g^{00}}\,\partial_{2}.\label{e2}
\end{equation}
Under such settings, 
\begin{equation}
T=\frac{1}{g^{00}\sqrt{-g}}\left[\partial_{1}(\sqrt{-g}\,g^{11}\,\partial_{1}+\partial_{2}(\sqrt{-g}\,g^{22}\,\partial_{2})\right]+\frac{m^{2}-\zeta R}{g^{00}}-\mathcal{Y}^{2},\label{e3}
\end{equation}
which would in turn imply that 
\begin{equation}
T=\frac{1}{g^{00}r}\left[-\partial_{r}\,r\,\partial_{r}-\partial_{\varphi}(\frac{r}{r^{2}}\,\partial_{\varphi})\right]+\frac{m^{2}-\zeta R}{g^{00}}-\mathcal{Y}^{2},\label{e4}
\end{equation}
In this case, 
\begin{equation}
H_{GFVT}=\tau_{3}\,\left(\frac{\mathcal{N}^{2}+T}{2\mathcal{N}}\right)+i\tau_{2}\,\left(\frac{-\mathcal{N}+T}{2N}\right)-i\,\mathcal{Y},\label{e5}
\end{equation}
where 
\begin{equation}
H_{GFVT}=\left(\begin{array}{c}
\phi_{1}\\
\phi_{2}
\end{array}\right)e^{-iEt}=i\,\partial_{t}\,\left(\begin{array}{c}
\phi_{1}\\
\phi_{2}
\end{array}\right)e^{-iEt}\label{e6}
\end{equation}
This would in turn yield 
\begin{equation}
(T+\mathcal{N}^{2})\,\phi_{1}+(T-\mathcal{N}^{2})\,\phi_{2}=(2i\mathcal{N}\mathcal{Y}+2\mathcal{N}E)\phi_{1},\label{e7}
\end{equation}
and 
\begin{equation}
-(T+\mathcal{N}^{2})\,\phi_{2}-(T-\mathcal{N}^{2})\,\phi_{1}=(2i\mathcal{N}\mathcal{Y}+2\mathcal{N}E)\phi_{2}.\label{e8}
\end{equation}
These equations lead to the following results 
\begin{equation}
\mathcal{N}(\phi_{1}-\phi_{2})=(E+i\mathcal{Y})(\phi_{1}+\phi_{2}),\label{e9}
\end{equation}
\begin{equation}
T(\phi_{1}+\phi_{2})=\mathcal{N}(E+i\mathcal{Y})(\phi_{1}-\phi_{2}).\label{e10}
\end{equation}
Obviously, Eq(\ref{e9}) suggests that if we take 
\[
\tilde{\psi}=\phi_{1}-\phi_{2}=\frac{1}{\mathcal{N}}(E+i\mathcal{Y})\psi
\]
 where 
\[
\psi=\phi_{1}+\phi_{2}
\]
then Eq. (\ref{e10}) would result 
\begin{equation}
T\psi=(E+i\mathcal{Y})^{2}\psi=(E^{2}+2iE\mathcal{Y}-\mathcal{Y}^{2})\psi.\label{e11}
\end{equation}
We may now substitute $T$ of (\ref{e4}) in (\ref{e11}) to obtain
\begin{equation}
\left[-\frac{1}{r}\partial_{r}\,r\,\partial_{r}-\frac{1}{r^{2}}\,\partial_{\varphi}^{2}+(m^{2}-\zeta R)\right]\psi=\left[g^{00}E^{2}+2iEg^{02}\partial_{\varphi}\right]\psi.\label{e12}
\end{equation}
We may now take 
\[
\psi\equiv\psi(r,\varphi)=\,e^{i\ell\varphi}R(r)/\sqrt{r}
\]
to obtain the 2-dimensional Schrödinger like KG equation 
\begin{equation}
\left[-\partial_{r}^{2}+\frac{(\ell^{2}-1/4)}{r^{2}}+\tilde{\Omega}^{2}r^{2}\right]R(r)=\lambda R(r),\label{e13}
\end{equation}
where
\[
\lambda=E^{2}-2\ell\Omega E-m^{2}-2\zeta\Omega^{2},\,\tilde{\Omega}^{2}=\Omega^{2}E^{2}
\]
and $R$ is the Ricci scalar curvature for the Gürses spacetime under
consideration is used, i.e., $R=g^{\mu\nu}R_{\mu\nu}=-2\Omega^{2}$.

It should be noted here that our FV-oscillators above are manifestations
of the very nature of Gürses spacetime in (\ref{eq:17}).

\subsection{PDM FV-oscillators in Gürses space-time}

In this subsection, we generalize FV-oscillators and include PDM FV-oscillators.
In so doing, one would use the non-minimal coupling form of the radial
momentum operator (i.e., $\hat{p}_{r}\rightarrow\hat{p}_{r}-i\mathcal{M}_{r}$
and recast Eq. (\ref{e12}) to accommodate such generalization as
\begin{equation}
\left[-\frac{1}{r}(\partial_{r}+\mathcal{M}_{r})\,r\,(\partial_{r}-\mathcal{M}_{r})-\frac{1}{r^{2}}\,\partial_{\varphi}^{2}+(m^{2}-\zeta R)\right]\psi=\left[g^{00}E^{2}+2iEg^{02}\partial_{\varphi}\right]\psi.\label{e14}
\end{equation}
This would allow us to write 
\begin{equation}
\left[-\partial_{r}^{2}+\frac{(\ell^{2}-1/4)}{r^{2}}+\tilde{\Omega}^{2}r^{2}+V_{PDM}(r)\right]R(r)=\lambda R(r),\label{e15}
\end{equation}
where 
\begin{equation}
V_{PDM}(r)=\frac{\mathcal{M}_{r}}{r}+\mathcal{M}_{r}^{\prime}+\mathcal{M}_{r}^{2};\,\,\mathcal{M}_{r}=\frac{f(r)'}{4f(r)}.\label{e16}
\end{equation}
Obviously, with $f(r)=1$ our PDM FV-oscillators would retrieve FV-oscillators
in (\ref{e13}).

Yet, with $f(r)=e^{2\eta r^{2}}\Rightarrow\mathcal{M}_{r}=\eta r$,
$V_{PDM}=\eta^{2}r^{2}+2\eta$, and hence 
\begin{equation}
\left[-\partial_{r}^{2}+\frac{(\ell^{2}-1/4)}{r^{2}}+\tilde{\gamma}^{2}r^{2}\right]R(r)=\tilde{\lambda}R(r),\label{e17}
\end{equation}
where $\tilde{\gamma}^{2}=\Omega^{2}E^{2}+\eta^{2}$ and 
\begin{equation}
\tilde{\lambda}=E^{2}-2\ell\Omega E-m^{2}-2\zeta\Omega^{2}-2\eta.\label{e17.1}
\end{equation}
It is clear that this equation is in the two-dimensional radial Schrödinger
oscillator form. Therefore, it admits an exact solution with eigenvalues
and radial eigenfunctions, respectively, given by
\begin{equation}
\tilde{\lambda}=2\left\vert \tilde{\gamma}\right\vert \left(2n_{r}+\left\vert \ell\right\vert +1\right)=2\left\vert \Omega E\right\vert \sqrt{1+\frac{\eta^{2}}{\Omega^{2}E^{2}}}\left(2n_{r}+\left\vert \ell\right\vert +1\right)\label{e18}
\end{equation}
and 
\begin{equation}
R\left(r\right)\sim r^{\left\vert \ell\right\vert +1/2}\exp\left(-\frac{\left\vert \tilde{\gamma}\right\vert r^{2}}{2}\right)L_{n_{n}}^{\left\vert \ell\right\vert }\left(\left\vert \tilde{\gamma}\right\vert r^{2}\right).\label{e19}
\end{equation}
Consequently, 
\begin{equation}
\psi\left(r\right)\sim r^{\left\vert \ell\right\vert }\exp\left(-\frac{\left\vert \tilde{\gamma}\right\vert r^{2}}{2}\right)L_{n_{n}}^{\left\vert \ell\right\vert }\left(\left\vert \tilde{\gamma}\right\vert r^{2}\right).\label{e20}
\end{equation}
Comparing (\ref{e17.1}) with (\ref{e18}) one obtains 
\begin{equation}
E^{2}-2\ell\Omega E-2\left\vert \Omega E\right\vert \sqrt{1+\frac{\eta^{2}}{\Omega^{2}E^{2}}}\left(2n_{r}+\left\vert \ell\right\vert +1\right)-\tilde{m}=0,\label{e21}
\end{equation}
where $\tilde{m}=m^{2}+2\zeta\Omega^{2}+2\eta$. This equation has
to be dealt with diligence since $|\Omega E|=+\Omega_{\pm}E_{\pm}$
or $|\Omega E|=-\Omega_{\mp}E_{\pm}$ , where $\Omega_{\pm}=\pm|\Omega|$
and $E_{\pm}=\pm|E|$. This would, in turn, manifestly suggest that
\begin{equation}
E_{\pm}^{2}-2\Omega_{\pm}E_{\pm}\left(\sqrt{1+\frac{\eta^{2}}{\Omega^{2}E^{2}}}\left[2n_{r}+\left\vert \ell\right\vert +1\right]+\ell\right)-\tilde{m}=0,\label{e22}
\end{equation}
for $|\Omega E|=+\Omega_{\pm}E_{\pm}$ , and 
\begin{equation}
E_{\pm}^{2}+2\Omega_{\mp}E_{\pm}\left(\sqrt{1+\frac{\eta^{2}}{\Omega^{2}E^{2}}}\left[2n_{r}+\left\vert \ell\right\vert +1\right]-\ell\right)-\tilde{m}=0,\label{e23}
\end{equation}
Eq. (\ref{e22}) would result 
\begin{equation}
E_{\pm}=\Omega_{\pm}k_{1}\pm\sqrt{\Omega^{2}k_{1}^{2}+\tilde{m}},\,\,k_{1}=\sqrt{1+\frac{\eta^{2}}{\Omega^{2}E^{2}}}\left[2n_{r}+\left\vert \ell\right\vert +1\right]+\ell,\label{e24}
\end{equation}
and Eq. (\ref{e23}) gives 
\begin{equation}
E_{\pm}=-\Omega_{\mp}k_{2}\pm\sqrt{\Omega^{2}k_{2}^{2}+\tilde{m}},\,\,k_{2}=\sqrt{1+\frac{\eta^{2}}{\Omega^{2}E^{2}}}\left[2n_{r}+\left\vert \ell\right\vert +1\right]-\ell.\label{e25}
\end{equation}
It is more convenient to report the energies in terms of positive,
$E_{\pm}^{(+)}$, and negative, $E_{\pm}^{(-)}$ velocities so that
for $\Omega=+|\Omega|$ we have 
\begin{equation}
E_{\pm}^{(+)}=\pm|\Omega|k_{\pm}\pm\sqrt{\Omega^{2}k_{\pm}^{2}+\tilde{m}},,\label{e26}
\end{equation}
and 
\begin{equation}
E_{\pm}^{(-)}=\pm|\Omega|k_{\mp}\pm\sqrt{\Omega^{2}k_{\mp}^{2}+\tilde{m}},\label{e27}
\end{equation}
where 
\begin{equation}
k_{\pm}=\sqrt{1+\frac{\eta^{2}}{\Omega^{2}E^{2}}}\left[2n_{r}+\left\vert \ell\right\vert +1\right]\pm\ell
\end{equation}
It is worth noting that the topic of spinless heavy particles in the
geometry formed by Gürses backgrounds has been studied in a number
of studies (for example,\citep{key-26,key-62,key-63}).

To obtain the FV form of the KG wave equation in curved manifolds,
we will use the approach provided in references \citep{key-65,key-66}.
The generalized Feshbach-Villars transformation is used (GFVT) An
identical transformation for characterizing both large and massless
particles was presented earlier in Ref.\citep{key-67}. The components
of the wave function $\Phi$ in the GFVT are provided by \citep{key-65}.
\begin{equation}
\psi=\phi_{1}+\phi_{2},\qquad i\tilde{\mathcal{D}}\psi=\mathcal{N}\left(\phi_{1}-\phi_{2}\right),\label{eq:19}
\end{equation}
where $\mathcal{N}$ is an arbitrary nonzero real parameter and $\tilde{\mathcal{D}}=\frac{\partial}{\partial t}+\mathcal{Y},$
is specified, with 
\begin{equation}
\mathcal{Y}=\frac{1}{2}\left\{ \partial_{i},\sqrt{-g}\frac{g^{0i}}{g^{00}}\right\} ,\label{eq:20}
\end{equation}
The anti-commutator is denoted by the curly bracket in Eq. \eqref{eq:20}.

Now, using the metric \eqref{eq:25}, it is simple to see that $R=0$,
implying that space-time is locally flat (no local gravity), and so
the coupling component is vanishing. The condition $\xi=0$ is known
as minimum coupling. However, in massless theory, $\xi$ equals 1/6.
(in 4 dimensions). The equations of motion are then conformally invariant
in this later instance.

A simple computation yields 
\[
\mathcal{Y}=\left(\frac{\sqrt{-g}}{g^{00}}\frac{\partial}{\partial\varphi}\right),
\]
 and we then obtain 
\begin{equation}
\mathcal{T}'=\mathcal{F}^{-1}\left(\frac{\partial}{\partial r}\left(-\sqrt{-g}\right)\frac{\partial}{\partial r}\right)\mathcal{F}^{-1}+\left(\left(\frac{\ell^{2}}{r^{2}}\sqrt{-g}\right)+\frac{m^{2}-2\varsigma\Omega^{2}}{g^{00}}-\left(\frac{\Omega\ell}{g^{00}}\sqrt{-g}\right)^{2}-E^{2}\right)\mathcal{F}^{-2},\label{eq:24}
\end{equation}
Using these techniques to obtain the Hamiltonian \eqref{eq:21}, one
may suppose a solution of the type 
\begin{equation}
\Phi(t,r,\varphi)=\Phi(r)e^{-i\left(Et-\ell\varphi\right)},\label{eq:25}
\end{equation}
where $\ell=0,\pm1,\pm2,..$. The KG equation \ref{eq:15} may be
written equivalently to the following two coupled equations 
\begin{align}
\left(\mathcal{N}^{2}+\mathcal{T}\right)\phi_{1}+\left(-\mathcal{N}^{2}+\mathcal{T}'\right)\phi_{2} & =2\mathcal{N}E\phi_{1}\nonumber \\
-\left(\mathcal{N}^{2}+\mathcal{T}\right)\phi_{2}-\left(-\mathcal{N}^{2}+\mathcal{T}'\right)\phi_{1} & =2\mathcal{N}E\phi_{2},\label{eq:26}
\end{align}
The sum and difference of the two previous equations yields a second
order differential equation for the field $\psi'=\mathcal{F}\psi$.
As a result, the radial equation is as follows: 
\begin{equation}
\left[\frac{d^{2}}{dr^{2}}+\frac{1}{r}\frac{d}{dr}-\frac{\zeta^{2}}{r^{2}}+\kappa\right]\psi\left(r\right)=0,\label{eq:27-1}
\end{equation}
where we have set 
\begin{equation}
\zeta=\ell,\qquad\kappa=\sqrt{E^{2}-m^{2}+2\varsigma\Omega^{2}}.\label{eq:28}
\end{equation}
We can observe that Eq. \eqref{eq:28} is a Bessel equation and its
general solution is defined by \citep{key-61} 
\begin{equation}
\psi\left(r\right)=A\,J_{|\zeta|}\left(\kappa r\right)+B\,Y_{|\zeta|}\left(\kappa r\right),\label{eq:29}
\end{equation}
where $J_{|\zeta|}\left(\kappa r\right)$ and $Y_{|\zeta|}\left(\kappa r\right)$
are the Bessel functions of order $\zeta$ and of the first and the
second kind, respectivement. Here $A$ and $B$ are arbitrary constants.
We notice that at the origin when $\zeta=0$, the function $J_{|\zeta|}\left(\kappa r\right)\ne0$.
However, $Y_{|\zeta|}\left(\kappa r\right)$ is always divergent at
the origin. In this case, we will consider only $J_{|\zeta|}\left(\kappa r\right)$
when $\zeta\ne0$. 

Hence, we write the solution to Eq. \eqref{eq:28} as follows 
\begin{equation}
\Psi\left(r\right)=\mathcal{A}\,J_{|\ell|}\left(\sqrt{E^{2}-m^{2}+2\varsigma\Omega^{2}}\,r\right),\label{eq:30}
\end{equation}
We can now express the whole two-component wavefunction of the spinless
heavy KG particle in the space-time of a Gürses using this solution.
\begin{equation}
\Psi\left(\boldsymbol{r}\right)=\left|\mathcal{A}_{1}\right|\left(\begin{array}{c}
1+\frac{E}{\mathcal{N}}\\
1-\frac{E}{\mathcal{N}}
\end{array}\right)e^{-i\left(Et-\ell\varphi\right)}\,J_{|\ell|}\left(\sqrt{E^{2}-m^{2}+2\varsigma\Omega^{2}}\,r\right),\label{eq:31}
\end{equation}
The constant $\left|\mathcal{A}_{1}\right|$ can be obtained by applying
the appropriate normalization condition to the KG equation (e.g.,
see Ref.\citep{key-69,key-70}), but it is fortunate that failing
to determine the normalization constants throughout this manuscript
has no effect on the final results.

Now we'll look at the specific instance where we wish to extend the
GFVT for the KGO. In general, we must substitute the momentum operator
in Eq. \eqref{eq:15}. As a result, Eq. \eqref{eq:30} may be rewritten
as follows. 
\begin{equation}
\mathcal{T}'=\mathcal{F}^{-1}\left(\frac{\partial}{\partial r}-m\omega r\right)\left(-\sqrt{-g}\right)\left(\frac{\partial}{\partial r}+m\omega r\right)\mathcal{F}^{-1}+\left(\left(\frac{\ell^{2}}{r^{2}}\sqrt{-g}\right)+\frac{m^{2}-2\varsigma\Omega^{2}}{g^{00}}-\left(\frac{\Omega\ell}{g^{00}}\sqrt{-g}\right)^{2}\right)\mathcal{F}^{-2},\label{eq:32}
\end{equation}
Similarly, the following differential equation may be obtained using
a simple calculation based on the approach described above. 
\begin{equation}
\left[\frac{d^{2}}{dr^{2}}+\frac{1}{r}\frac{d}{dr}-m^{2}\omega^{2}r^{2}-\frac{\sigma^{2}}{r^{2}}+\delta\right]\psi\left(r\right)=0,\label{eq:33}
\end{equation}
with 
\begin{equation}
\sigma^{2}=\left(\ell\right)^{2},\qquad\delta=E^{2}-m^{2}+2\varsigma\Omega^{2}+\left(\Omega\ell\right)^{2}+2m\omega.\label{eq:34}
\end{equation}
The KGO for a spin-0 particle in the space-time of a static cosmic
string is given by Eq. \eqref{eq:33}. To derive the solution to this
problem, we first suggest a radial coordinate transformation
\begin{equation}
\mathcal{\gamma}=m\omega r^{2},\label{eq:35}
\end{equation}
Subsitutuing the expression for $\chi$ into Eq. \eqref{eq:33}, we
obtain 
\begin{equation}
\left[\frac{d^{2}}{d\mathcal{\mathcal{\gamma}}^{2}}+\frac{1}{\mathcal{\mathcal{\gamma}}}\frac{\partial}{d\mathcal{\mathcal{\gamma}}}-\frac{\sigma^{2}}{4\mathcal{\mathcal{\gamma}}^{2}}+\frac{\delta}{4m\omega\mathcal{\gamma}}-\frac{1}{4}\right]\psi\left(\mathcal{\gamma}\right)=0.\label{eq:36}
\end{equation}
So, if we look at the asymptotic behavior of the wave function at
the origin and infinity, and we're looking for regular solutions,
we may assume a solution of the type 
\begin{equation}
\psi\left(\mathcal{\gamma}\right)=\mathcal{\mathcal{\gamma}}^{\frac{\left|\sigma\right|}{2}}e^{-\frac{\mathcal{\mathcal{\gamma}}}{2}}F\left(\mathcal{\mathcal{\gamma}}\right),\label{eq:37}
\end{equation}
As previously, we can plug this back into Eq. \eqref{eq:36}, and
we get 
\begin{equation}
\mathcal{\mathcal{\gamma}}\frac{d^{2}F\left(\mathcal{\mathcal{\gamma}}\right)}{d\mathcal{\mathcal{\gamma}}^{2}}+\left(|\sigma|+1-\chi\right)\frac{dF\left(\mathcal{\mathcal{\gamma}}\right)}{d\mathcal{\gamma}}-\left(\frac{|\sigma|}{2}-\frac{\delta}{4m\omega}+\frac{1}{2}\right)F\left(\mathcal{\mathcal{\gamma}}\right)=0,\label{eq:38}
\end{equation}
This is the confluent hypergeometric equation \citep{key-64}, the
solutions to which are defined in terms of the kind of confluent hypergeometric
function. 
\begin{equation}
F\left(\mathcal{\gamma}\right)=_{1}F_{1}\left(\frac{\left|\sigma\right|}{2}-\frac{\delta}{4m\omega}+\frac{1}{2},|\sigma|+1,\mathcal{\gamma}\right),\label{eq:39}
\end{equation}
We should note that the solution \eqref{eq:39} must be a polynomial
function of degree $n$. However, taking $n\rightarrow\infty$ imposes
a divergence issue. We can have a finite polynomial only if the factor
of the last term in Eq. \eqref{eq:38} is a negative integer, meaning,
\begin{equation}
\frac{\left|\sigma\right|}{2}-\frac{\delta}{4m\omega}+\frac{1}{2}=-n\qquad,n=0,1,2,\cdots.\label{eq:40}
\end{equation}
With this result and the parameters \eqref{eq:34}, we may derive
the quantized energy spectrum of KGO in the Gürses space-time, and
hence, 
\begin{equation}
E^{\pm}\left(n\right)=\pm\sqrt{4m\omega n+2m\omega\left|\ell\right|+m^{2}-2\varsigma\Omega^{2}-\left(\Omega\ell\right)^{2}},\label{eq:41}
\end{equation}
We may notice that the energy relies clearly on the angular deficit
$\alpha$. In other words, because to the presence of the wedge angle,
the curvature of space-time that is impacted by the topological defect,
i.e., the cosmic string, would affect the relativistic dynamics of
the scalar particle by creating a gravitational field.

The corresponding wave function is given by 
\begin{equation}
\Psi\left(\boldsymbol{r}\right)=\left|\mathcal{A}_{2}\right|\left(m\omega r^{2}\right)^{\frac{\left|\ell\right|}{2}}e^{-\frac{m\omega r^{2}}{2}}{}_{1}F_{1}\left(\frac{\left|\ell\right|}{2}-\frac{\delta}{4m\omega}+\frac{1}{2},\left|\ell\right|+1,m\omega r^{2}\right),\label{eq:42}
\end{equation}
Thereafter, the general eigenfunctions are written as 
\begin{equation}
\Psi\left(\boldsymbol{r}\right)=\left|\mathcal{A}_{2}\right|\left(\begin{array}{c}
1+\frac{E}{\mathcal{N}}\\
1-\frac{E}{\mathcal{N}}
\end{array}\right)\left(m\omega r^{2}\right)^{\frac{\left|\ell\right|}{2}}e^{-\frac{m\omega r^{2}}{2}}e^{-i\left(Et-\ell\varphi\right)}{}_{1}F_{1}\left(\frac{\left|\ell\right|}{2}-\frac{\delta}{4m\omega}+\frac{1}{2},\left|\ell\right|+1,m\omega r^{2}\right),\label{eq:43}
\end{equation}
where $\left|\mathcal{A}_{2}\right|$ is the normalization constant.

\section{the Feshbach-Villars oscillator under Potentials In the Gürses space-time,}

This section focuses on the Klein-Gordon oscillator (KGO) within the
context of a cosmic displacement. By examining the Geodesic Flow Vector
Theory (GFVT) as demonstrated in Section 2, one can obtain the equations
of motion for a scalar particle. Various authors have investigated
the quantum dynamics of relativistic particles in the space-time of
a cosmic displacement, and several models have been explored. For
instance, in a prior publication, Mazur studied the quantum mechanical
properties of heavy or massless particles in the gravitational field
of a rotating cosmic displacement \Citep{key-72}. He showed that
energy would be quantized when the string has non-zero rotational
momentum. Additionally, Gerbert and Jackiw \Citep{key-73} demonstrated
solutions to the KG and Dirac equations in the (2+1)-dimensional space-time
created by a large point particle with arbitrary angular momentum.
The vacuum expectation value of the stress-energy tensor for a massless
scalar field conformally related to gravity was studied in Ref. \Citep{key-74}.
The authors of \Citep{key-73} investigated the behavior of a quantum
test particle subjected to the Klein-Gordon equation in a space-time
created by a rotating cosmic string.

Furthermore, in a different study (Ref. \Citep{key-75}), it was shown
that the extrema of the energy of the field, given specific angular
and linear momenta, can be described as solutions of spinning cosmic
strings in $U(1)$ scalar field theory, which have an energy density
that is cylindrically symmetric. Additionally, Ref. \Citep{key-76}
investigated topological and geometrical phases arising from the gravitational
field of a cosmic string with mass and angular momentum.

Researchers investigating the dynamics and properties of relativistic
quantum particles have recently become interested in the gravitational
effects of rotating cosmic strings. For example, in Ref. \Citep{key-77},
the vacuum fluctuations surrounding a spinning cosmic string were
examined for a massless scalar field using a renormalization approach.
Similarly, Ref. \citep{key-78} investigated the vacuum polarization
of a scalar field in the gravitational field of a rotating cosmic
string. In addition, the authors of Ref. \Citep{key-80} used a fully
relativistic approach to study the Landau levels of a heavy spinless
particle in the space-time created by a rotating cosmic string.

In the study by Wang et al.\Citep{key-81}, the Klein-Gordon oscillator
(KGO) was investigated in the presence of a homogeneous magnetic field
within the context of a rotating cosmic string. Similarly, in Ref.
\Citep{key-81} , the dynamics of a spinless relativistic particle
subject to a uniform magnetic field in the space-time created by a
rotating cosmic string were examined. Additionally, Ref. \Citep{key-82}
reported on the relativistic quantum dynamics of a Klein-Gordon scalar
field exposed to a Cornell potential in the space-time of a rotating
cosmic string. Moreover, Ref. \Citep{key-83} studied the relativistic
behavior of a scalar charged particle in the space-time of a rotating
cosmic string with a Cornell-type potential and the Aharonov-Bohm
effect.

\subsection{Free Feshbach-Villars equation in the Gürses space-time under a Coulomb-type
potential. }

In this section, we expand on the topic discussed in Sec 2 to include
a more general Gürses space-time with non-zero angular momentum. Specifically,
we examine the behavior of a heavy, relativistic spin-0 particle with
a wave function represented by Psi that satisfies the Klein-Gordon
equation \ref{eq:15} in the space-time created by a (2+1)-dimensional
stationary Gürses with Coulomb-type potentials $\left(S\left(r\right)=\frac{\lambda}{r}\right)$.

\begin{equation}
\left[\frac{d^{2}}{dr^{2}}+\frac{1}{r}\frac{d}{dr}-\frac{\ell^{2}}{r^{2}}+\left(E-S\left(r\right)\right)^{2}-m^{2}+2\varsigma\Omega^{2}+\left(\Omega\ell\right)^{2}\right]\varphi\left(r\right)=0,\label{eq:44}
\end{equation}
where 
\begin{equation}
\left[\frac{d^{2}}{dr^{2}}+\frac{1}{r}\frac{d}{dr}-\frac{\ell^{2}}{r^{2}}+\left(E-\frac{\lambda}{r}\right)^{2}-m^{2}+2\varsigma\Omega^{2}+\left(\Omega\ell\right)^{2}\right]\varphi\left(r\right)=0,\label{eq:45}
\end{equation}
After simple algebraic of equation \eqref{eq:45}: 
\begin{equation}
\left[\frac{d^{2}}{dr^{2}}+\frac{1}{r}\frac{d}{dr}-\frac{\left(\ell^{2}-\lambda^{2}\right)}{r^{2}}-2\frac{E\lambda}{r}+E^{2}-m^{2}+2\varsigma\Omega^{2}+\left(\Omega\ell\right)^{2}\right]\varphi\left(r\right)=0,\label{eq:46}
\end{equation}
After simple algebraic manipulations we arrive at the following second
order differential equation for the radial function $\varphi(r)$
\begin{equation}
\varphi\left(r\right)=\left|\mathcal{A}_{3}\right|r^{-\frac{1}{2}}\mathrm{WhittakerM}\left[-\frac{E\lambda}{\sqrt{m^{2}-2\varsigma\Omega^{2}-\Omega^{2}\ell^{2}-E^{2}}},\sqrt{\ell^{2}-\lambda^{2}},2r\left(\sqrt{m^{2}-2\varsigma\Omega^{2}-\Omega^{2}\ell^{2}-E^{2}}\right)\right],\label{eq:47}
\end{equation}
By simplifying the relationship between Whitakar and confluent hypergeometric
function\citep{key-61}, yields{\scriptsize{} 
\begin{equation}
\varphi\left(r\right)=\left|\mathcal{A}_{3}\right|r^{-\frac{1}{2}}\mathrm{e}^{-\left(\sqrt{-\Omega^{2}\ell^{2}-E^{2}+m^{2}-2\varsigma\Omega^{2}}\right)r}\left(2\sqrt{-\Omega^{2}\ell^{2}-E^{2}+m^{2}-2\varsigma\Omega^{2}}\,r\right){}^{\frac{1}{2}+\sqrt{\ell^{2}-\lambda^{2}}}g\left(r\right)\label{eq:48}
\end{equation}
}where{\scriptsize{} 
\begin{equation}
g\left(r\right)=1F1\left(\frac{2\sqrt{\ell^{2}-\lambda^{2}}\left(\sqrt{-\Omega^{2}\ell^{2}-E^{2}+m^{2}-2\varsigma\Omega^{2}}+1\right)+2E\lambda}{2\sqrt{-\Omega^{2}\ell^{2}-E^{2}+m^{2}}-2\varsigma\Omega^{2}},1+2\sqrt{\ell^{2}-\lambda^{2}},2r\left(\sqrt{-\Omega^{2}\ell^{2}-E^{2}+m^{2}-2\varsigma\Omega^{2}}\right)\right)\label{eq:49}
\end{equation}
}We should note that the solution must be a polynomial function of
degree $n$. However, taking $n\rightarrow\infty$ imposes a divergence
issue. We can have a finite polynomial only if the factor of the last
term in Eq. is a negative integer, meaning, 
\begin{equation}
\frac{2\sqrt{\ell^{2}-\lambda^{2}}\left(\sqrt{-\Omega^{2}\ell^{2}-E^{2}+m^{2}-2\varsigma\Omega^{2}}+1\right)+2E\lambda}{2\sqrt{-\Omega^{2}\ell^{2}-E^{2}+m^{2}-2\varsigma\Omega^{2}}}=-n\qquad,n=0,1,2,\cdots.\label{eq:50}
\end{equation}
With this result and the parameters \eqref{eq:34}, we may derive
the quantized energy spectrum of KGO in the Gürses space-time, and
hence,{\scriptsize{} 
\begin{equation}
E^{\pm}(n)=\pm\frac{\left(2\left(\sqrt{\ell^{2}-\lambda^{2}}\right)+2n+1\right)\left[\sqrt{\left(8n\left(\sqrt{\ell^{2}-\lambda^{2}}\right)+4\left(\sqrt{\ell^{2}-\lambda^{2}}\right)+4\left(\ell^{2}+n^{2}+n+\frac{1}{4}\right)\right)\left(m^{2}-2\varsigma\Omega^{2}-\Omega^{2}\ell^{2}\right)}\right]}{\left(8n+4\right)\left(\sqrt{\ell^{2}-\lambda^{2}}\right)+4\left(\ell^{2}+n^{2}+n+\frac{1}{4}\right)},\label{eq:51}
\end{equation}
}where $\left|\mathcal{C}_{3}\right|$ is an integration constant.
The complete eigenstates are given by{\scriptsize{} 
\begin{equation}
\Psi\left(\boldsymbol{r}\right)=\left|\mathcal{A}_{3}\right|\left(\begin{array}{c}
1+\frac{E}{\mathcal{N}}\\
1-\frac{E}{\mathcal{N}}
\end{array}\right)e^{-i\left(Et-\ell\varphi\right)}r^{-\frac{1}{2}}\mathrm{e}^{-\left(\sqrt{-\Omega^{2}\ell^{2}-E^{2}+m^{2}-2\varsigma\Omega^{2}}\right)r}\left(2\sqrt{-\Omega^{2}\ell^{2}-E^{2}+m^{2}-2\varsigma\Omega^{2}}\,r\right){}^{\frac{1}{2}+\left(\sqrt{\ell^{2}-\lambda^{2}}\right)}g\left(r\right)\label{eq:52}
\end{equation}
}{\scriptsize\par}

\subsection{Feshbach-Villars oscillator in the Gürses space-time under a Coulomb-type
potential. }

From now on we proceed to study the Feshbach--Villars Oscillator
in a Gürses space-time. Firstly, we start by considering a scalar
quantum particle embedded in the background gravitational field of
the space-time described by the metric \eqref{eq:42}. In this way,
we shall introduce a replacement of the momentum operator $p_{i}\longrightarrow p_{i}+im\omega x_{i}$
where $p_{i}=i\nabla_{i}$ in Eq. \eqref{eq:57}. Then, we have 
\begin{equation}
\left[\frac{d^{2}}{dr^{2}}+\frac{1}{r}\frac{d}{dr}-m^{2}\omega^{2}r^{2}-\left(\frac{\ell^{2}-\lambda^{2}}{r^{2}}\right)-\frac{2\lambda E}{r}+E^{2}-m^{2}-2\varsigma\Omega^{2}+\left(\Omega\ell\right)^{2}+2mw\right]\varphi\left(r\right)=0,\label{eq:53}
\end{equation}
Based on the prior studies, we will apply the GFVT to the case of
KGO in the relevant space using the same techniques as previously.
Inserting Eqs. \eqref{eq:52} and \eqref{eq:45} into the Hamiltonian
\eqref{eq:56}, then assuming the solution \eqref{eq:48}, yields
two linked differential equations comparable to Eq.\eqref{eq:56},
but with different values of $\mathcal{T}^{\prime}$.

Manipulating exactly the same steps before, we obtain the following
radial equation 
\begin{equation}
\left[\frac{\partial^{2}}{\partial r^{2}}+\frac{1}{r}\frac{\partial}{\partial r}-m^{2}\omega^{2}r^{2}-\frac{\vartheta^{2}}{r^{2}}-\frac{2\lambda E}{r}+\delta\right]\varphi\left(r\right)=0,\label{eq:54}
\end{equation}
where we have defined 
\begin{equation}
\vartheta^{2}=\ell^{2}-\lambda^{2},\qquad\beta^{2}=E^{2}-m^{2}+2\varsigma\Omega^{2}+2mw+\left(\Omega\ell\right)^{2}.\label{eq:55}
\end{equation}
Let us now conside $\mathcal{P}=\sqrt{m\omega}r$, and therefore rewrite
the radial equation (\eqref{eq:53}) as follows ($\varphi\left(r\right)\rightarrow\varphi\left(\mathcal{\mathcal{P}}\right)$):
\begin{equation}
\left[\frac{d^{2}}{d\mathcal{\mathcal{P}}^{2}}+\frac{1}{\mathcal{\mathcal{P}}}\frac{d}{d\mathcal{\mathcal{P}}}-\frac{\vartheta^{2}}{\mathcal{P}^{2}}-\frac{\delta}{\mathcal{P}}-\mathcal{P}^{2}+\frac{\beta^{2}}{m\omega}\right]\varphi\left(\mathcal{\mathcal{P}}\right)=0,\label{eq:56}
\end{equation}
where we have defined a new parameter 
\begin{equation}
\delta=2\lambda\left(m\omega^{-1}\right)^{\frac{1}{2}},\label{eq:57}
\end{equation}
Let us look at the asymptotic behavior of the solutions to Eq. (\eqref{eq:55}),
which are found for $\mathcal{\mathcal{P}}\rightarrow0$ and$\mathcal{\mathcal{P}}\rightarrow\infty$.
The behavior of the potential solutions to Eq. (\eqref{eq:55}) at
$\mathcal{\mathcal{P}}\rightarrow0$ and allows us to define the function
$R(\mathcal{P})$ in terms of an unknown function $\mathcal{O}(\mathcal{P})$
as follows from Refs. \citep{key-51,key-52,key-53}: 
\begin{equation}
\varphi(\mathcal{P})=\mathcal{P}^{\left|\gamma\right|}e^{-\frac{\mathcal{P}^{2}}{2}}\mathcal{O}\left(\mathcal{P}\right),\label{eq:58}
\end{equation}
Hence, by plugging the radial wave function from Eq. (11) into Eq.
(9), we get 
\begin{equation}
\frac{d^{2}\mathcal{O}\left(\mathcal{P}\right)}{d\mathcal{\mathcal{P}}^{2}}+\left[\frac{\left(2\left|\gamma\right|+1\right)}{\mathcal{\mathcal{P}}}-2\mathcal{P}\right]\frac{d\mathcal{O}\left(\mathcal{P}\right)}{d\mathcal{\mathcal{P}}}+\left[\frac{\beta^{2}}{m\omega}-2\left(2\left|\gamma\right|+1\right)-\frac{\delta}{\mathcal{P}}\right]\mathcal{O}\left(\mathcal{P}\right)=0,\label{eq:59}
\end{equation}
The Heun biconfluent equation \citep{key-51,key-55,key-56,key-57}
relates to the second order differential equation (\eqref{eq:58}),
and the function $\mathcal{O}\left(\mathcal{K}\right)$ is the Heun
biconfluent function. 
\begin{equation}
\mathcal{O}\left(\mathcal{P}\right)=HeunB\left(2\left|\gamma\right|,0,\frac{\beta^{2}}{m\omega},2\delta,\mathcal{P}\right)\label{eq:60}
\end{equation}
To continue our discussion of bound state solutions, let us employ
the Frobenius method\citep{key-88,key-91}. As a result, the solution
to Equation (\eqref{eq:59}) may be expressed as a power series expansion
around the origin: 
\begin{equation}
\mathcal{O}\left(\mathcal{P}\right)=\stackrel[s=0]{\infty}{\sum}a_{s}\mathcal{P}^{s}\label{eq:61}
\end{equation}
We find the following recurrence connection by substituting the series
(\eqref{eq:60}) into (\eqref{eq:59}): 
\begin{equation}
a_{s+2}=\frac{\delta}{\left(s+2\right)\left(s+1+\zeta\right)}a_{s+1}-\frac{\left(\varTheta-2s\right)}{\left(s+2\right)\left(s+1+\zeta\right)}a_{s}\label{eq:62}
\end{equation}
where $\zeta=2\left|\gamma\right|+1$ and $\varTheta=\frac{\beta^{2}}{m\omega}-2\left(\left|\gamma\right|+1\right)$.
We may determine the additional coefficients of the power series expansion
by starting with $a_{0}=1$ and applying the relation (\eqref{eq:60}).
(\eqref{eq:58}). As an example, 
\begin{equation}
a_{1}=\frac{\delta}{\zeta}=2\sqrt{\frac{m}{\omega}}\lambda\left(\frac{1}{2\left|\gamma\right|+1}\right)\label{eq:63}
\end{equation}
\begin{align}
a_{2}= & \frac{\delta^{2}}{2\zeta\left(1+\zeta\right)}-\frac{\varTheta}{2\left(1+\zeta\right)}\nonumber \\
= & \left(\frac{m}{\omega}\right)\lambda^{2}\frac{1}{\left(2\left|\gamma\right|+1\right)\left(\left|\gamma\right|+1\right)}-\frac{\varTheta}{4\left(\left|\gamma\right|+1\right)}\label{eq:64}
\end{align}
The wave function must be normalizable, as is widely known in quantum
theory. As a result, we suppose that the function $\mathcal{O}\left(\mathcal{P}\right)$
disappears at $\mathcal{\mathcal{P}}\rightarrow0$ and $\mathcal{\mathcal{P}}\rightarrow\infty$.
This indicates that we have a finite wave function everywhere, which
means that there is no divergence of the wave function at $\mathcal{\mathcal{P}}\rightarrow0$
and $\mathcal{\mathcal{P}}\rightarrow\infty$ thus bound state solutions
may be produced.

In Eq \eqref{eq:58}, we have, however, expressed the function $\mathcal{O}\left(\mathcal{P}\right)$
as a power series expansion around the origin (64). By demanding that
the power series expansion \eqref{eq:60} or the Heun biconfluent
series becomes a polynomial of degree $n$, bound state solutions
can be obtained. As a result, we ensure that $\mathcal{O}\left(\mathcal{P}\right)$
behaves as $\mathcal{P^{\left|\gamma\right|}}$ at $\mathcal{P}\rightarrow0$
the origin and disappears at\citep{key-84,key-85} . We can see from
the recurrence relation \eqref{eq:64} that by applying two constraints
\citep{key-83,key-84,key-85,key-86,key-87,key-88,key-89,key-90,key-91,key-92},
the power series expansion \eqref{eq:60} becomes a polynomial of
degree $n$: 
\begin{equation}
\varTheta=2n\,\,and\,\,a_{n+1}=0\label{eq:65}
\end{equation}
where $n=1,2,3,....$ From the condition $\varTheta=2n$, we can obtain:
\begin{equation}
E^{\pm}\left(n\right)=\pm\sqrt{2m\left(n+\left|\gamma\right|\right)+m^{2}-2\varsigma\Omega^{2}-2\varsigma\Omega^{2}}\label{eq:66}
\end{equation}
The corresponding wave function is given by 
\begin{equation}
\varphi\left(r\right)=\left|\mathcal{A}_{4}\right|\mathcal{P}^{\left|\gamma\right|}e^{-\frac{\mathcal{P}^{2}}{2}}HeunB\left(2\left|\gamma\right|,0,\frac{\beta^{2}}{m\omega},2\delta,\mathcal{P}\right)\label{eq:67}
\end{equation}
the final expression of the wave-function of the spinless FVO propagating
in the Gürses background can be represented as 
\begin{equation}
\Psi\left(\boldsymbol{r}\right)=\left|\mathcal{A}_{4}\right|\left(\begin{array}{c}
1+\frac{E}{\mathcal{N}}\\
1-\frac{E}{\mathcal{N}}
\end{array}\right)e^{-i\left(Et-\ell\varphi\right)}\mathcal{P}^{\left|\gamma\right|}e^{-\frac{\mathcal{P}^{2}}{2}}HeunB\left(2\left|\gamma\right|,0,\frac{\beta^{2}}{m\omega},2\delta,\mathcal{P}\right),\label{eq:68}
\end{equation}
where the parameters $\vartheta$ and $\delta$ are defined in Eq.\eqref{eq:55}.

\section{Conclusion }

The objective of this research is to investigate the relativistic
behavior of spinless quantum particles by applying the Feshbach-Villars
approach to two models: one involving the interaction between KGO
and the gravitational field generated by the background geometry of
Gürses space-time and the PDM of FV in Gürses space-time, and the
other involving Gürses space-time with a Coulomb-like potential. To
achieve this, the FV formulation of scalar fields in Minkowski space-time
was modified to create comparable formulations on two different curved
manifolds. The study provides exact solutions for both systems and
presents energy spectra that depend on the parameters that define
the space-time topology. The wave-functions of the quantum system
are expressed as confluent hypergeometric functions for the free Feshbach-Villars
equation in Gürses space-time, which is not unexpected since the former
can be described throughout the latter using a suitable coordinate
transformation.

\end{document}